\documentclass[11pt, twocolumn]{article}
\usepackage{amsmath,amsfonts,amsthm,amssymb}
\usepackage{graphicx,float,wrapfig}
\usepackage{fullpage}
\usepackage{listings}
\usepackage{color}
\usepackage{wrapfig}
\usepackage[authoryear, round]{natbib}
\usepackage{abstract}
\usepackage{ccaption}
\usepackage{algorithmic}
 
\captionnamefont{\bfseries}
\captiontitlefont{\itshape}
\captiondelim{: }
\captionwidth{0.45\textwidth}
\changecaptionwidth

\title{Spatio-temporal Patterns in Multi-Electrode Array Local Field Potential Recordings}
\author{Bronwyn Woods}
\date{June 2011}

\begin{document}

\twocolumn[
\maketitle
\begin{onecolabstract}
This paper presents a method for the detection of traveling waves of activity in neural recordings from multielectrode arrays. The method converts local field potential measurements into the phase domain and fits a series of linear models to find planar traveling waves of activity. Here I present the new approach in the context of the previous work it extends, apply the approach to data from neural recordings from a single animal, and verify the success of the method on simulated data. 
\end{onecolabstract}
]

\section{Background and Overview}

Current technology allows for neural recording on many scales, each with trade offs in terms of spatial resolution, temporal resolution, and recording target. A single extracellular electrode has high temporal resolution and relatively high spatial precision.   Electrode arrays, by recording from spatially structured collections of electrodes simultaneously, supplement this with some information on spatio-temporal patterns.  Electrode arrays may therefore be a useful recording technique for exploring local network dynamics \textit{in vivo}.

The signal recorded by a single electrode contains high frequency electrical signals from nearby spiking neurons and a low frequency signal referred to as the local field potential (LFP).  The LFP is believed to be a measure of the summed synaptic activity in a radius around an electrode \citep{Katzner2009}. Previous work in the Hatsopoulos lab at the University of Chicago \citep{Rubino06} suggests that LFP recorded over an array of electrodes can reveal spatio-temporal patterns in activity. Specifically, it can reveal traveling waves of oscillation.
  
Traveling waves are widely discussed in the neuroscience literature \citep{Delaney1994, Huang2004, Manjarrez07, Wu2008, Xu2007, Reimer10}. Propagating waves of oscillation have been reported in the visual, auditory, olfactory and somatosensory systems as well as in motor cortex, hippocampus, and spontaneous cortical activity.  Theoretical work reported in \citet{Ermentrout2001} demonstrates that the traveling wave patterns reported in experimental studies are likely to arise in realistic neural networks.  Even so, the possible computational purpose, if any, of traveling waves remains unclear.  Determining their significance, or even relegating them to the realm of the epiphenomenal, will require advances in the methodology used to detect and describe them.

Currently, there is no generally established methodology for detecting and describing traveling waves.  Many studies rely on visual inspection of Voltage Sensitive Dye (VSD) images.  Several recent papers using LFP electrode recordings make a start on developing formal wave detection techniques.  \citet{Lubenov2009} used regression methods on a collection of electrodes, and \citet{Rubino06} used phase derivative calculations on electrode array data.  This project synthesizes ideas from these two papers (applied to the data from the later) to develop methodology that allows for statistically rigorous single trial analysis of LFP array data.

This research analyzes data from the Hatsopoulos Lab at the University of Chicago.  I first replicated the results reported in \citet{Rubino06} regarding the detection of traveling waves.  This previous methodology was able to detect the presence of wavelike activity and capture its broad characteristics, but was unable to provide a single trial description of activity.  I made several improvements to the methodology, allowing for single trial analysis.  I also demonstrated that wavelike activity is persistent rather than occurring in short isolated time periods as previously reported.  Finally, I created artificial data to test both the original and improved methodology under varying noise conditions, demonstrating the efficacy of my improvements.

The next two sections of this paper are introductory. Section \ref{sec:waves} states the model for a planar wave of oscillation.  Section \ref{sec:data} describes the data being modeled.  The following section discusses the results of previous analysis, while section \ref{sec:analysis} describes my analysis and results in detail.  Finally, section \ref{sec:artificial} describes the conclusions from artificial data analysis.

\section{Traveling Waves}
\label{sec:waves}

A traveling wave is composed of a collection of oscillators whose relative phases depend on their spatial location.  The simplest form of this phase relationship is a planar wave, where the phase of the oscillators is linear in space and time.  Without loss of generality, we can set the phase at time 0 and position 0 to be 0. In one dimension, the equation to describe a planar wave can then be written
\begin{equation}
V(x,t) \propto sin\left[ 2\pi  \left( ft - \frac{f}{s}x\right)\right]
\end{equation}
where $f$ is the frequency of the oscillators and $s$ is the speed of the wave.

This linear phase relationship can be easily extended to describe planar waves in two dimensions by making $x$ a vector.  With two spatial dimensions, more complex patters are possible such as bull's eye or spiral waves.  However, this research addresses only the simpler, linear case.

Since the phase is linear, fitting this model requires estimating three parameters: the change in phase in the $x$, $y$ and $t$ dimensions.  From these parameters we can calculate properties of the wave that are of scientific interest such as its speed and direction.  Table \ref{wavepars} shows this relationship.

\renewcommand{\arraystretch}{2}
\captionwidth{0.8\textwidth}
\begin{table*}
	\begin{centering}
	\begin{tabular}{ | l | p{10cm} | c | }
		\hline 
		\bf{Parameter} & \bf{Interpretation} & \bf{Calculation} \\ \hline \hline 
		Frequency & The frequency of the oscillators composing the wave. The current model requires all oscillators to have the same frequency. & $\hat{f} \propto \hat{\delta_t}$\\ \hline
		Direction & Wave propagation direction. & $\hat{d} \propto atan\left( \frac{\hat{\delta_y}}{\hat{\delta_x}}\right)$ \\ \hline
		Speed & Spatial speed of contours of constant phase.  Higher speeds indicate more synchronous activity. & $\frac{\hat{\delta_t}}{\sqrt{\hat{\delta_x}^2 + \hat{\delta_y}^2}}$ \\
 		\hline
	\end{tabular}
	\caption{Calculating wave parameters from linear model parameters.  Here $\delta_x$, $\delta_y$, and $\delta_t$ are the change in phase in the $x$, $y$, and $t$ dimensions. Wave speed and direction are the primary parameters of scientific interest.}
	\label{wavepars}
\end{centering}
\end{table*}
\captionwidth{0.45\textwidth}

\section{Data}
\label{sec:data}

The data I analyze here were recorded by Nicho Hatsopoulos' lab at the
University of Chicago.  They consist of 391 trials of a center-out reaching task performed
by a monkey.  On each trial, the monkey was given a target
in one of  eight directions (the ``instruction'') and then had to wait for one second before initiating
movement to the target (after the ``go'' cue).  

\begin{figure*}
\begin{centering}
\includegraphics[width=0.8\textwidth]{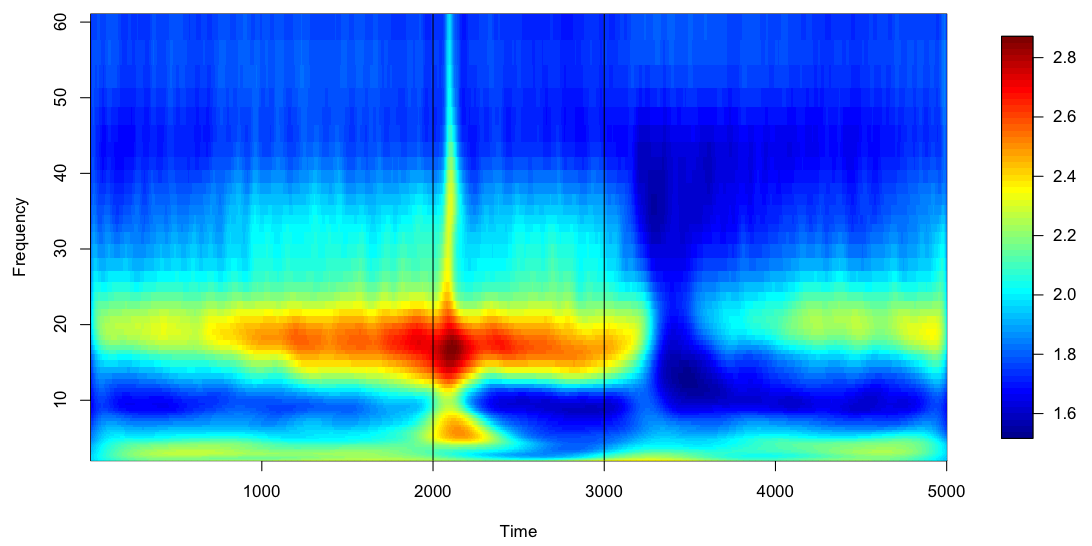}
\caption{This spectrogram, computed using Morlet wavelets, shows high power between 15 and 20 Hz during the waiting period (between 2000 and 3000 ms in this figure).  The spectrogram was computed individually for each trial and then averaged over all trails, aligning on the time at which the monkey received the direction cue.}
\label{waveletwaiting}
\end{centering}
\end{figure*}

LFP voltages were recorded on a 96
channel Utah array implanted in motor cortex.  The data were sampled at 1 kHz. Based on prior research \citep{Rubino06} I chose the waiting period between the ``instruction'' and ``go'' cues to be the period of interest for further analysis.  As shown in Figure \ref{waveletwaiting}, this period is characterized by strong oscillatory activity in the Beta band between 15 and 20 Hz.

\captionwidth{0.8\textwidth}
\captionwidth{0.45\textwidth}

Fourier analysis of the waiting period data revealed the expected peak in the Beta band.  Additionally, the Fourier spectrum revealed significant high frequency noise at multiples of 60 Hz, especially in two channels.  This noise is presumably electrical line noise, and is removed by filtering during analysis. 

%

\begin{figure*}
 \begin{centering}
 \includegraphics[width=0.8\textwidth]{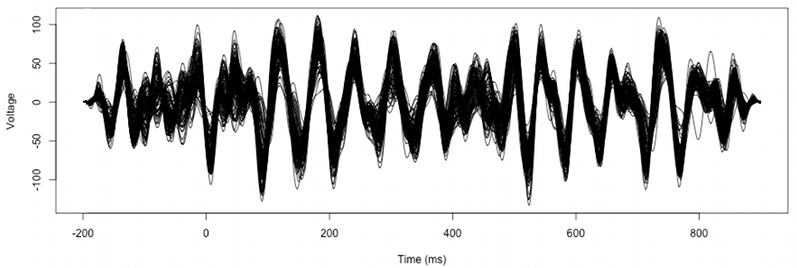}
 \caption{Voltage from all channels superimposed for one trial.  The data were wide-band filtered between 10 and 45 Hz before being plotted.}
 \label{widevoltage}
 \end{centering}
 \end{figure*}

The 96 channels on the array are very highly correlated.  Figure \ref{widevoltage} shows the voltages for all channels superimposed for one trial.  For this figure, the data were wide-band filtered between 10 and 45 Hz for clarity, but the correlation of the channels and strong oscillatory activity around 20 Hz is quite clear.

\captionwidth{0.85\textwidth}
\captionwidth{0.45\textwidth}

\section{Summary of Previous Work}
\label{sec:prevwork}

Previous work reported in \citet{Rubino06} found evidence of wavelike activity.  The directions of wave propagation were seen to follow a bimodal distribution with peaks approximately $\pi$ radians apart (nearly opposite directions).  The wavelike activity was found to consist of short periods of waves surrounded by non-wavelike periods.  Here I give a very brief summary of these results and the analysis used to reach them.  More details can be found in the Analysis section or in the original paper.

As described in section \ref{sec:waves}, fitting a planar wave model to data requires estimating three parameters: $\delta_x$, $\delta_y$, and $\delta_t$ (the change in phase in each of the three dimensions).  The first step of the analysis is to extract the phase.  This is done using the analytic signal, discussed in detail in section \ref{sec:phase}.  At each point in space and time, the local gradient of the phase is computed.  These gradients are then averaged spatially to give estimates $\hat{\delta_x}$, $\hat{\delta_y}$, and $\hat{\delta_t}$ at each time point.

\begin{figure}
 \begin{centering}
 \includegraphics[width=0.5\textwidth]{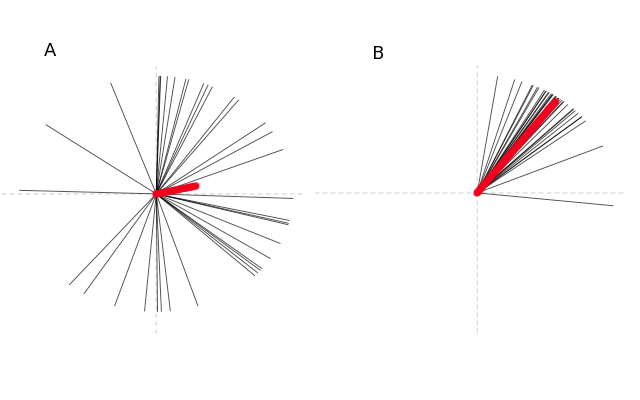}
 \caption{Two time points demonstrating high and low phase gradient alignment.  Each black line indicates the direction of the spatial gradient on one of the channels of the array at the time point.  The red line is the mean over the array.  The length of the red line is the MRL, a measure of phase gradient alignment.  At time 1 the gradients are well aligned (indicating wavelike activity).  At time 2, the gradients are not aligned (indicating non wavelike activity)}
 \label{phasealign}
 \end{centering}
 \end{figure}

In addition to fitting the three parameters, it is necessary to evaluate the model fit and determine whether activity is indeed wavelike.  Previous work used a measure of phase gradient alignment.  If the data are actually a planar wave, then the phase will be linear in space and time and the spatial phase gradients at all locations will point in the same direction.  If the data are not wavelike, the spatial gradients would be unlikely to align.  Figure \ref{phasealign} demonstrates this contrast.

To measure phase alignment, Rubino et al. used a measure they called Phase Gradient Directionality (PGD).  
\begin{equation}
	PGD =  \frac{||\frac{1}{n}\sum_{i=1}^{n}{(\delta_{xi}, \delta_{yi})}||}{\frac{1}{n}\sum_{i=1}^{n}{||(\delta_{xi}, \delta_{yi})||}}
\end{equation}
where $n$ ranges over electrodes.  This is very similar to the standard circular variance measure known as Mean Resultant Length or MRL \citep{Mardia99}.  
\begin{equation}
	MRL = || \frac{1}{n} \sum_{i=1}^{n}{\frac{(\delta_{xi}, \delta_{yi})}{||(\delta_{xi}, \delta_{yi}) ||}} ||
\end{equation}
The only difference is that PGD weights individual gradients according to their magnitude while MRL normalizes all gradients.  The two measures are very similar for practical purposes. 

\begin{figure*}
 \begin{centering}
 \includegraphics[width=\textwidth]{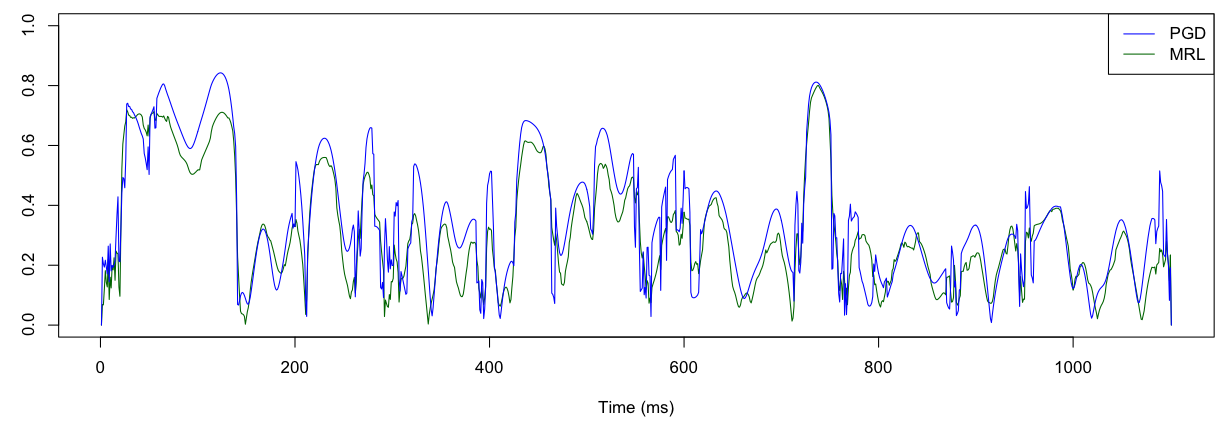}
 \caption{Alignment measures computed for an example trial.  Previous work identifies waves and time periods when the alignment measure is above 0.5.  This leads to the identification of short periods of wavelike activity, such as those highlighted in the figure.  The alignment measures appear quite noisy, making single trial interpretation questionable.}
 \label{aligntrial}
 \end{centering}
 \end{figure*}

The alignment measure is computed for each time point.  Since wavelike activity should have high alignment, Rubino et al. set a threshold of 0.5 and classified times at which the PGD was above this value to be wavelike.  The alignment measure for an example trial is shown in figure \ref{aligntrial}.

\captionwidth{0.8\textwidth}
\captionwidth{0.45\textwidth}

\begin{figure}
 \begin{centering}
 \includegraphics[width=0.5\textwidth]{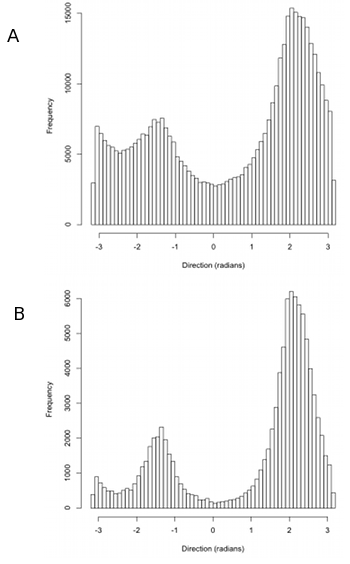}
 \caption{Histograms of estimated wave propagation direction.  (A) shows estimated directions over all times and trials.  (B) shows estimated directions for those times identified as being wavelike.  The bimodal pattern of propagation directions is clear, especially in B.}
 \label{directionhist}
 \end{centering}
 \end{figure}

Once the model is fit to the data, it is possible to examine the wave characteristics such as propagation direction.  Figure \ref{directionhist} shows histograms of wave propagation direction for all times and for times classified as waves.  Especially during the wavelike periods, the propagation directions show a clear bimodal pattern.

\begin{figure*}
 \begin{centering}
 \includegraphics[width=\textwidth]{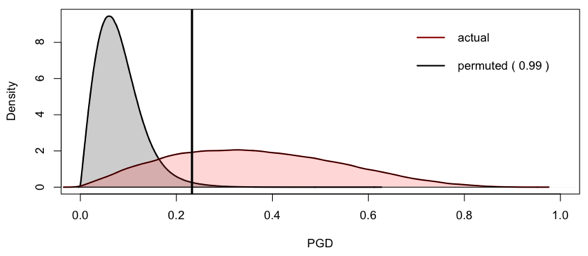}
 \caption{The results of a permutation procedure to derive a null distribution for the alignment measure (PGD) used in previous work.  The grey distribution is the null distribution, and the black line indicates the 99th percentile of the null distribution.  This indicates that an appropriate threshold for the PGD would be 0.2 rather than 0.5.}
 \label{permutetest}
 \end{centering}
 \end{figure*}

The conclusions from previous work were that the data show short periods of wavelike activity during which the waves propagate in two primary directions.  I performed a permutation test to determine if the PGD threshold of 0.5 was appropriate for wave identification.  By permuting the spatial organization of the channels I destroyed any spatial wavelike pattern while maintaining the temporal structure of the individual channels and the overall level of correlation across the array.  In this way I computed a null distribution for the alignment measures, shown in figure \ref{permutetest}.  

\captionwidth{0.8\textwidth}
\captionwidth{0.45\textwidth}

Comparing the true PGD distribution to this null distribution, I concluded that the threshold of 0.5 was inappropriate.  A better threshold would be closer to 0.2.  This indicates that the wavelike activity in the data is a more pervasive and consistent pattern in the system than previous work suggested.  The rest of this paper describes the analysis methods I used to extract interpretable single trial results showing nearly constant wavelike activity.

\section{Improved Analysis Methods}
\label{sec:analysis}

A planar traveling wave of oscillation is defined by linear dependence
of phase on spatio-temporal position.  For this reason, a natural approach to
detecting such waves is to work in the phase domain.  This approach is
taken in papers such as \citet{Rubino06} and \citet{Lubenov2009}.  The first step of this modeling approach is to extract the phase at each point in space and time.  This process is described in section \ref{sec:phase}.  Once the data is transformed into the phase domain, the remaining analysis involves fitting a linear model and evaluating model fit.
 

\subsection{Extracting Phase}
\label{sec:phase}

The first step of a phase-based analysis is to extract the instantaneous
phase at each spatio-temporal point.  This is typically
accomplished using the Hilbert transform to calculate the analytic
signal, as described below.

The Fourier transform of a real-valued function is Hermitian.  Because of this, the information
about negative frequencies is redundant.  The analytic signal discards
these negative frequency components at the cost of converting the
associated signal from real-valued to complex-valued.  If $x(t)$ is a real valued
function with Fourier spectrum $X(f)$, the corresponding analytic
signal is defined to be

\begin{equation}
\label{analytic}
x_a(t) = x(t) + i\hat{x}(t)
\end{equation}

where $\hat{x}(t)$ is the Hilbert transform of $x(t)$.  The Hilbert
transform is the convolution of $x(t)$ with $\frac{1}{\pi t}$.  We can
therefore see that equation \eqref{analytic} becomes

\begin{align}
x_a(t) &= x(t) + i\left(x(t) * \frac{1}{\pi t}\right)\\
&= x(t) * \left(\delta(t)+ \frac{i}{\pi t}\right)
\end{align}

Taking the Fourier transform of this, and defining $h(f)$ to be the
Heaviside step function, we see that the negative
frequencies are 0, as expected.

\begin{align}
X_a(f) &= X(f) \cdot 2h(f)\\
&= \begin{cases}
2X(f) & f>0\\
X(f) & f=0\\
0 & f<0
\end{cases}
\end{align}

\begin{figure}
\begin{centering}
\includegraphics[width=0.5\textwidth]{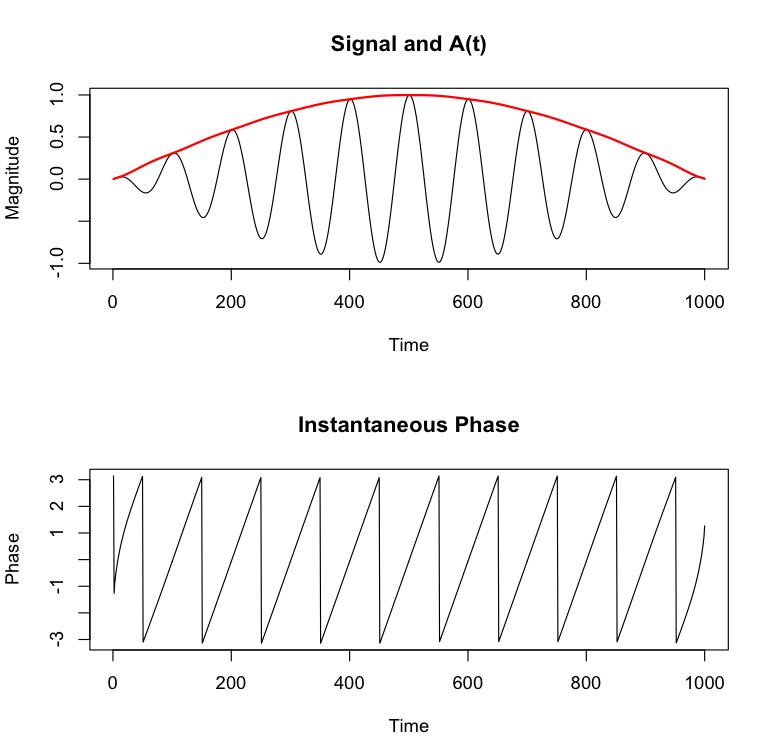}
\caption{The instantaneous amplitude and phase calculated via the analytic signal for an amplitude modulated pure sine wave.  The upper graph shows the original signal in black with the amplitude envelope $A(t)$ plotted in red.  The lower graph shows the instantaneous wrapped phase $\phi (t)$.}
\label{analytic_summary}
\end{centering}
\end{figure}

Returning to the time domain analytic signal,  we can write equation
\eqref{analytic} in polar coordinates as

\begin{equation}
x_a(t) = A(t)e^{i\phi (t)}
\end{equation}

where $A(t)$ is called the amplitude envelope and $\phi (t)$ is the
instantaneous phase, wrapped between $-\pi$ and $\pi$.  Figure \ref{analytic_summary} shows the instantaneous amplitude and phase described by the analytic signal for an amplitude modulated pure sine wave.

One concern in using the analytic signal to extract phase is that $\phi(t)$ is only interpretable as instantaneous phase for signals with narrow band frequency content. 
%
The data in this research is not a pure sine wave, so this characteristic of the analytic signal is a potential problem.  In addition, the model for a planar wave has only one frequency term, requiring a single frequency for all oscillators.  However, since the data do show a well-defined peak frequency, it is reasonable to filter the data and look for wavelike activity around that peak.  Previous work used a wide-band filter from 10-45Hz.  My work shows that the previous filter was too wide.  Interpretable results require a much narrower band filter; the results reported here used a filter with a 1 Hz passband (plus rolloff).

%

\subsection{Unwrapping Phase}
\label{sec:unwrap}

As shown in the previous section, the phase as extracted by the analytic signal is wrapped between $-\pi$ and $\pi$. Fitting a model to the data requires removing the $2\pi$ jumps in the phase.  This is a well-studied problem known as ``phase unwrapping''.  With data satisfying strong assumptions about the sampling rate and noise levels, phase unwrapping is a trivial problem.  However, with realistic data, the problem is unidentifiable.  The multitudes of unwrapping algorithms that have been proposed aim to find a good solution in the realistic case.

To unambiguously unwrap the phase, we must make the assumption that the data were sampled according to the Nyquist-Shannon sampling theorem in all dimensions; that is, that the data were sampled at a minimum rate of twice the highest frequency present in the data.  If this assumption is satisfied, the true phase difference between any adjacent points will be less that $\pi$.  Therefore, unwrapping can be done by traversing the data with some spanning tree, adding or subtracting units of $2\pi$ to a point if its value differs from that of the previous point by more than $\pi$.

When local noise is introduced into the data, it may cause violations of the assumption.  The basic unwrapping algorithm may erroneously add or subtract $2\pi$ or fail to do so when it should.  Since the unwrapping is done along some tree, any points that are unwrapped subsequent to the error along the same branch of the tree will have the same error.  This error propagation can result in wildly inaccurate unwrapping results.  The general approach of any improved unwrapping algorithm is to chose the structure of the unwrapping tree such that noisy areas are unwrapped last and error is not propagated. 

Previous work used very simple phase unwrapping, repeatedly unwrapping small portions of the data in one dimension at a time.  As described in section \ref{sec:prevwork}, \citet{Rubino06} estimated the model parameters by computing the local phase gradient at each point and then averaging over the array.  Computing each component of the gradient only requires using several data points in a line in a single dimension and so it is only strictly necessary to unwrap those points.  Rubino et al. unwrap the necessary data points, compute the gradient component, discard the unwrapping results and then repeat the process for each dimension at each point.  This unwrapping technique restricts dramatic error propagation in the unwrapping by only unwrapping small portions of data at a time.  However, it does nothing to limit propagation of error within the local regions and restricts possible model fitting techniques to those that do not require simultaneous consideration of a large set of data points (precluding, for instance, regression approaches).

It can be demonstrated that considering all available dimensions during unwrapping can yield better results.  This is because unwrapping requires choosing a path or tree to traverse the data.  In one dimension, the path is predetermined. In higher dimensions, there are increasing numbers of paths to choose from, increasing the likelihood that one can be found that avoids traveling through noisy regions of data.  There are dozens of unwrapping algorithms that have been proposed for two dimensions \citep{Ghiglia98}.  At least two types of unwrapping algorithms have been extended to three dimensions: branch cut algorithms \citep{Salfity06, Huntley2001, Marklund2007} and quality guided unwrapping \citep{Cusack2002, Abdul-Rahman2007, AbdulRahman09}.  Branch cut algorithms identify singularities in the phase volume and use them to identify and remove unwrapping paths that will lead to error.  The phase is then unwrapped using a breadth-first spanning tree. Quality guided algorithms use some heuristic to evaluate the quality of each possible connection between adjacent points.  The highest quality edges are unwrapped first.  Examples of these quality heuristics are local variance of phase values or of phase differences. This research involved a three dimensional phase volume, so I implemented these 3-D unwrapping algorithms and applied them to the Utah array data.

We know from examining the voltage data that the LFP is highly correlated over the array channels.  Since all the channels are oscillating together, we observe that the phase should differ by no more than $2\pi$ across the array at any given time.  This observation gives us a heuristic for evaluating the performance of phase unwrapping algorithms: calculating the mean phase difference across the array for the unwrapped phase.  Though the performance of the algorithms varies somewhat with the passband chosen, the quality guided algorithms generally outperformed the branch cut algorithm. For the rest of the analysis, the data were unwrapped with the quality guided unwrapping algorithm using the local variance of phase differences as the quality heuristic.  


\subsection{Fitting the Model}

The planar wave model is linear.  Therefore, modeling the unwrapped phase using the established technique of linear regression is appropriate.  The model is the simple linear model
\begin{equation}
	\phi(x,y,t) = \beta_0 + \beta_1x + \beta_2y + \beta_3t.
\end{equation}
However, we must be able to account for changes over time so we cannot fit the entire phase volume simultaneously.  Instead, we fit multiple separate models to overlapping time windows of data.  To estimate the wave parameters at time $t$ using a window size of 5ms, we use data from time $[t-2]$ms to $[t+2]$ms. This means that modeling a 2 second section of data requires fitting 1,996 separate models, one for each time point from 3ms to 1,998ms.

Once all the models are fitted, we can estimate the wave parameters along with standard confidence intervals.  We can also examine the $R^2$ values as one indicator of whether the planar wave model is a good fit to the data.  Finally, we can perform formal hypothesis tests to determine whether there is any statistically significant linear dependence of phase on spatial position.  If we fail to reject the null hypothesis that $\beta_1 = \beta_2 = 0$, we can conclude that we have no evidence for wavelike activity.

\begin{figure*}
\begin{centering}
\includegraphics[width=\textwidth]{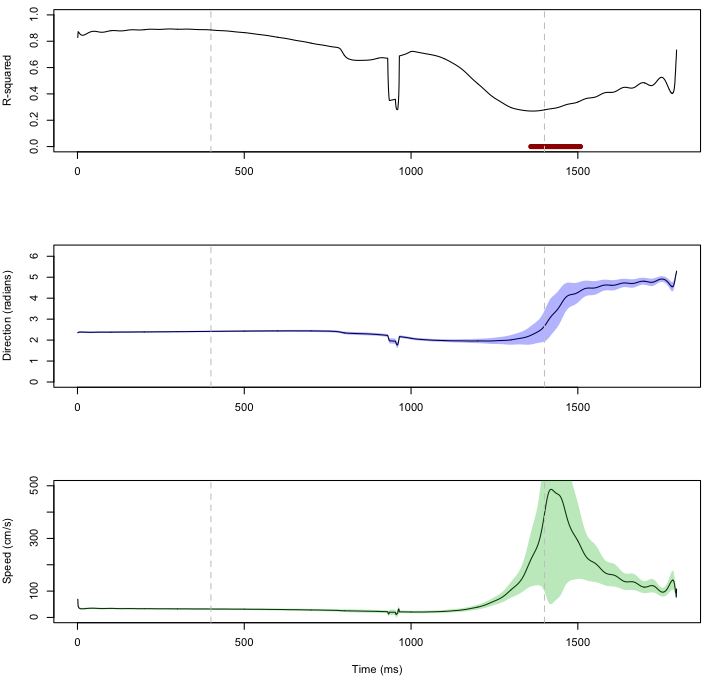}
\caption{Results from analyzing a single trial of data in the 17-18Hz band.  For each plot, time runs along the horizontal axis and the values for each time point are calculated using the linear regression centered at that time point.  The top plot shows the $R^2$ values.  The middle plot shows estimated wave propagation direction.  The bottom plot shows estimated wave speed.  The shaded areas are 95\% confidence intervals.  Vertical grey lines indicate the beginning and ending of the waiting period.  Though the switch in direction in this trial occurs near the end of the waiting period, there is no clear pattern of this type apparent over all trials.}
\label{regresults}
\end{centering}
\end{figure*}

By modeling narrow-band filtered data using 3D phase unwrapping and linear regression we can produce interpretable estimates of wave parameters on a single trial basis.  Figure \ref{regresults} shows the results of this analysis for one trial.  In this trial we see two primary wave propagation directions and a switch between them.  For most of the trial, the speed is around 40 cm/s, which is physiologically reasonable \citep{Rubino06}.  At the time when the wavelike activity switches direction, the speed increasing drastically as the LFP activity becomes temporarily more synchronous.  Additionally, during this transition we fail to reject the null hypothesis of the test described above.  Overall, the null hypothesis was rejected for over 98\% of the time points analyzed in the 17-18Hz band.  This indicates that the wavelike activity is a highly persistent feature of the pre-motor neural system.

\captionwidth{0.85\textwidth}
\captionwidth{0.45\textwidth}

\subsection{Evaluating the Model}

\begin{figure*}
\begin{centering}
\includegraphics[width=0.9\textwidth]{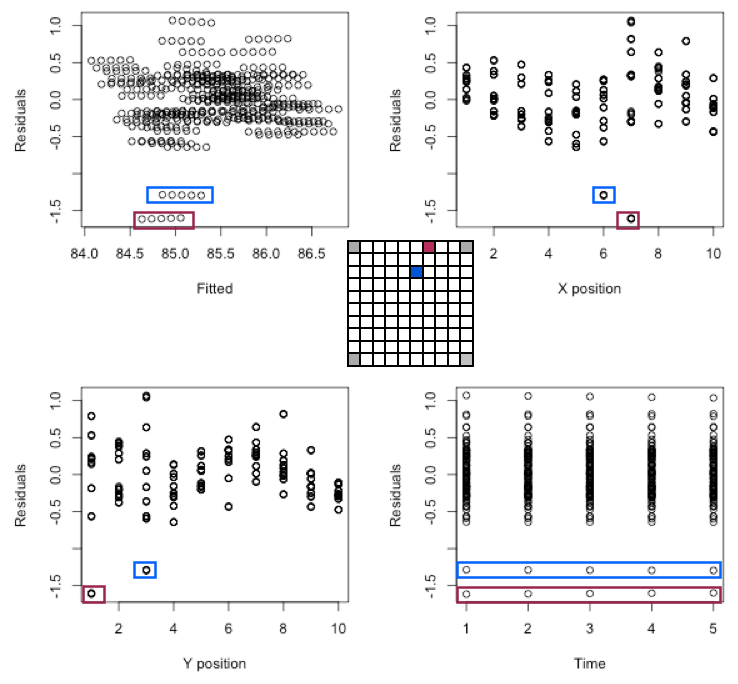}
\caption{Residuals plotted against model terms for a regression with low $R^2$.  The highlighted residuals are outlying data points from two channels.  The schematic in the center shows the location of these channels.}
\label{outliers}
\end{centering}
\end{figure*}

The linear regression framework allows us to use standard analysis techniques to evaluate model performance.  For instance, identifying outliers that result from isolated noise or unwrapping errors can explain some decreases in $R^2$ values such as that observed at around time 950 in figure \ref{regresults}. Figure \ref{outliers} shows the residuals for one of the regressions fit to data in this time period.  The residuals reveal outliers limited to two channels.  Identifying and removing outlying channels could allow improved model fitting.

\captionwidth{0.85\textwidth}
\captionwidth{0.45\textwidth}

\section{Artificial Data}
\label{sec:artificial}

\begin{figure}
\begin{centering}
\includegraphics[width=0.5\textwidth]{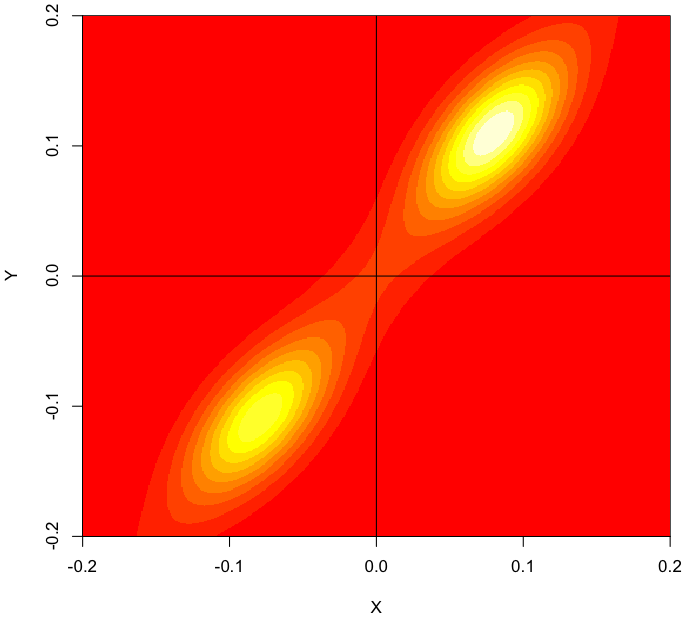}
\caption{The distribution over $\delta_x$ and $\delta_y$ used to generate artificial data.}
\label{artsample}
\end{centering}
\end{figure}

\begin{figure}
\begin{centering}
\includegraphics[width=0.5\textwidth]{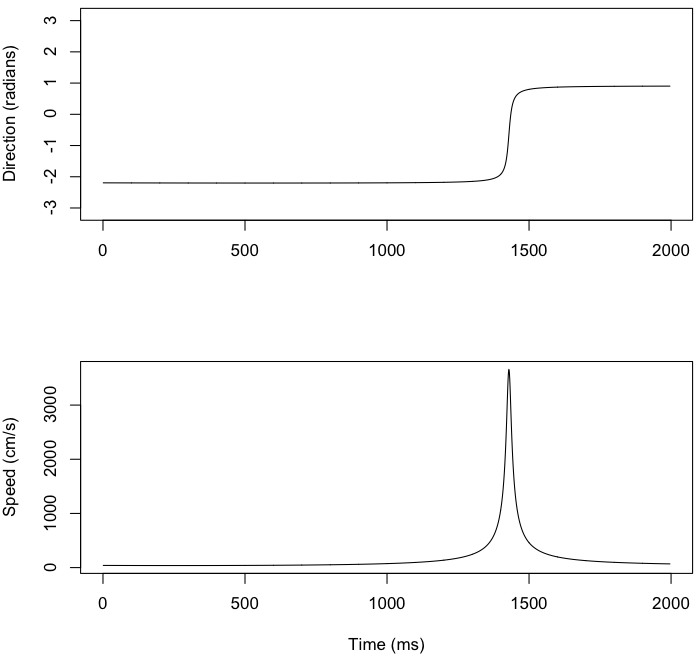}
\caption{An example of the direction and speed values for an artificial data trial.}
\label{arttrial}
\end{centering}
\end{figure}

In order to assess the improved and previous analysis techniques, we created artificial data that mimicked the bimodal wave direction distribution observed in the real data.  The artificial data were generated using model parameters generated using temporally smoothed Metropolis-Hastings sampling from a bimodal distribution.  This distribution is shown in figure \ref{artsample}.  The frequency term of the artificial data model is specified to be 17.5Hz. The wave parameters of an example trial of artificial data is shown in figure \ref{arttrial}.  By adding various types of noise to this artificial data, we can evaluate the performance of wave detection methodology.

When no noise is added to the artificial data, both the original and improved methodologies do a good job of estimating both the speed and direction of the waves.  Both methods make some significant error on less than 0.5\% of the data during times when the speed and direction of the artificial waves are changing rapidly.

\begin{figure}
\begin{centering}
\includegraphics[width=0.5\textwidth]{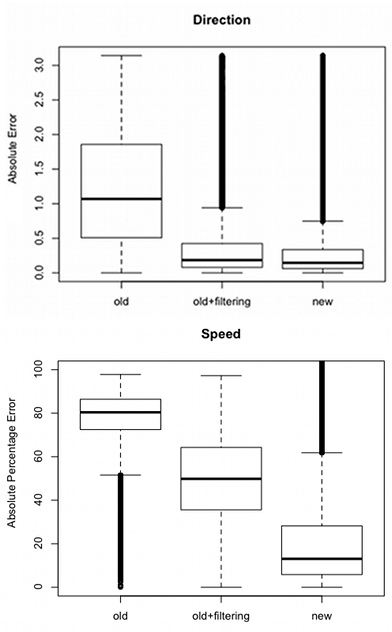}
\caption{Mean error measures for estimation of artificial data parameters. ``Old'' indicates the original method, ``old+filtering'' uses the original estimation technique with the new filtering technique, and ``new'' uses the entire improved processing pipeline.  The apparently heavy tails on the box plots represent less than half a percent of the data but appear heavy because each box plot is generated using around 800,000 points.}
\label{artboxplot}
\end{centering}
\end{figure}

When noise is added to the system, differences between the methods become apparent.  Figure \ref{artboxplot} shows the mean estimation errors for three different methods when filtered white noise is added to the artificial data.  The three methods are (1) the original method from \citet{Rubino06}, (2) the original estimation technique using the new narrow band filtering, and (3) the new method.  The estimation errors from the three models show that the narrow band filtering is very important to both speed and direction estimation while the new model fitting technique further improves estimation of speed.

\section{Conclusions and Future Work}

The work reported in this paper makes a new scientific claim about the pattern of wavelike activity observed in monkey motor cortex: it is a persistent pattern, characteristic of the system.  This contrasts with previous work which suggested that the wavelike activity was episodic.  The methodology developed in this research allows single trial analysis of the data, and simulated data confirms that it results in improved estimation of wave parameters.  This improvement in analysis methodology opens the door for additional investigation of the properties and purpose of traveling waves in the neural system.

The largest estimation improvements resulted from modifications of the filtering procedure used before extracting instantaneous phase.  Because of the characteristics of the Hilbert transform and analytic signal, it is very important to apply a narrow band filter.  A principled method for selecting the appropriate width and location of the pass band for filtering is a matter for future investigation.  It is possible to repeat the analysis for multiple narrow frequency bands, thereby covering a wide frequency space.  However, methodology for integrating the results from multiple frequency bands has not yet been developed.  It is scientific question whether is is appropriate to treat LFP data as a truly narrow band signal.  The current planar wave model has only a single term for frequency, so if we wish to model data with multiple frequency components the model would need to be extended.

Phase-based modeling of wavelike activity is the most common approach, but it would also be possible to develop a method for modeling the LFP voltage directly.  Since the voltage model for a traveling wave is nonlinear, fitting such a model would require nonlinear estimation techniques.  However, modeling the voltage would avoid the phase extraction and phase unwrapping process which introduces strict filtering requirements and possible unwrapping error.  If the planar wave model cannot be fit successfully to data, other voltage domain techniques are possible.  \citet{Gabriel2003} present a method based on pairwise correlations between channels, successfully demonstrating wavelike activity in monkey striate cortex using a linear 15-electrode array.

In summary, the work presented here continues the task of creating an automated, statistically principled detection methodology for the detection of planar waves in neural systems.  This could aid the community of neuroscientists attempting to determine the prevalence and functional significance of such spatio-temporal patterns.

\bibliographystyle{apa}
\bibliography{library}

\begin{thebibliography}{}

\bibitem[\protect\astroncite{Abdul-Rahman et~al.}{2009}]{AbdulRahman09}
Abdul-Rahman, H., Arevalillo-Herr\'{a}ez, M., Gdeisat, M., Burton, D., Lalor,
  M., Lilley, F., Moore, C., Sheltraw, D., and Qudeisat, M. (2009).
\newblock {Robust three-dimensional best-path phase-unwrapping algorithm that
  avoids singularity loops}.
\newblock {\em Applied Optics}.

\bibitem[\protect\astroncite{Abdul-Rahman et~al.}{2007}]{Abdul-Rahman2007}
Abdul-Rahman, H.~S., Gdeisat, M.~A., Burton, D.~R., Lalor, M.~J., Lilley, F.,
  and Moore, C.~J. (2007).
\newblock {Fast and robust three-dimensional best path phase unwrapping
  algorithm.}
\newblock {\em Applied optics}, 46(26):6623--35.

\bibitem[\protect\astroncite{Cusack and Papadakis}{2002}]{Cusack2002}
Cusack, R. and Papadakis, N. (2002).
\newblock {New Robust 3-D Phase Unwrapping Algorithms: Application to Magnetic
  Field Mapping and Undistorting Echoplanar Images*1}.
\newblock {\em NeuroImage}, 16(3):754--764.

\bibitem[\protect\astroncite{Delaney et~al.}{1994}]{Delaney1994}
Delaney, K.~R., Gelperin, a., Fee, M.~S., Flores, J.~a., Gervais, R., Tank,
  D.~W., and Kleinfeld, D. (1994).
\newblock {Waves and stimulus-modulated dynamics in an oscillating olfactory
  network.}
\newblock {\em Proceedings of the National Academy of Sciences of the United
  States of America}, 91(2):669--73.

\bibitem[\protect\astroncite{Ermentrout and Kleinfeld}{2001}]{Ermentrout2001}
Ermentrout, G.~B. and Kleinfeld, D. (2001).
\newblock {Traveling electrical waves in cortex: insights from phase dynamics
  and speculation on a computational role.}
\newblock {\em Neuron}, 29(1):33--44.

\bibitem[\protect\astroncite{Gabriel and Eckhorn}{2003}]{Gabriel2003}
Gabriel, A. and Eckhorn, R. (2003).
\newblock {A multi-channel correlation method detects traveling $\gamma$-waves
  in monkey visual cortex}.
\newblock {\em Journal of Neuroscience Methods}, 131(1-2):171--184.

\bibitem[\protect\astroncite{Ghiglia and Pritt}{1998}]{Ghiglia98}
Ghiglia, D.~C. and Pritt, M.~D. (1998).
\newblock {\em {Two-Dimensional Phase Unwrapping: Theory, Algorithms, and
  Software}}.
\newblock Wiley-Interscience.

\bibitem[\protect\astroncite{Huang et~al.}{2004}]{Huang2004}
Huang, X., Troy, W.~C., Yang, Q., Ma, H., Laing, C.~R., Schiff, S.~J., and Wu,
  J.-Y. (2004).
\newblock {Spiral waves in disinhibited mammalian neocortex.}
\newblock {\em The Journal of neuroscience : the official journal of the
  Society for Neuroscience}, 24(44):9897--902.

\bibitem[\protect\astroncite{Huntley}{2001}]{Huntley2001}
Huntley, J.~M. (2001).
\newblock {Three-dimensional noise-immune phase unwrapping algorithm.}
\newblock {\em Applied optics}, 40(23):3901--8.

\bibitem[\protect\astroncite{Katzner et~al.}{2009}]{Katzner2009}
Katzner, S., Nauhaus, I., Benucci, A., Bonin, V., Ringach, D.~L., and
  Carandini, M. (2009).
\newblock {Local origin of field potentials in visual cortex.}
\newblock {\em Neuron}, 61(1):35--41.

\bibitem[\protect\astroncite{Lubenov and Siapas}{2009}]{Lubenov2009}
Lubenov, E.~V. and Siapas, A.~G. (2009).
\newblock {Hippocampal theta oscillations are travelling waves.}
\newblock {\em Nature}, 459(7246):534--9.

\bibitem[\protect\astroncite{Manjarrez et~al.}{2007}]{Manjarrez07}
Manjarrez, E., Vazquez, M., and Flores, A. (2007).
\newblock {Computing the center of mass for traveling alpha waves in the human
  brain}.
\newblock {\em Brain Research}.

\bibitem[\protect\astroncite{Mardia and Jupp}{1999}]{Mardia99}
Mardia, K.~V. and Jupp, P.~E. (1999).
\newblock {\em {Directional Statistics}}.
\newblock Wiley.

\bibitem[\protect\astroncite{Marklund et~al.}{2007}]{Marklund2007}
Marklund, O., Huntley, J.~M., and Cusack, R. (2007).
\newblock {Robust unwrapping algorithm for three-dimensional phase volumes of
  arbitrary shape containing knotted phase singularity loops}.
\newblock {\em Optical Engineering}, 46(8):085601.

\bibitem[\protect\astroncite{Reimer et~al.}{2010}]{Reimer10}
Reimer, A., Hubka, P., Engel, A., and Kral, A. (2010).
\newblock {Fast Propagating Waves within the Rodent Auditory Cortex}.
\newblock {\em Cerebral Cortex}.

\bibitem[\protect\astroncite{Rubino et~al.}{2006}]{Rubino06}
Rubino, D., Robbins, K.~A., and Hatsopoulos, N.~G. (2006).
\newblock {Propagating waves mediate information transfer in the motor cortex}.
\newblock {\em Nature Neuroscience}.

\bibitem[\protect\astroncite{Salfity et~al.}{2006}]{Salfity06}
Salfity, M.~F., Ruiz, P.~D., Huntley, J.~M., Graves, M.~J., Cusack, R., and
  Beauregard, D.~A. (2006).
\newblock {Branch cut surface placement for unwrapping of undersampled
  three-dimensional phase data: application to magnetic resonance imaging
  arterial flow mapping}.
\newblock {\em Applied Optics}.

\bibitem[\protect\astroncite{Wu et~al.}{2008}]{Wu2008}
Wu, J.-Y., {Xiaoying Huang}, and {Chuan Zhang} (2008).
\newblock {Propagating waves of activity in the neocortex: what they are, what
  they do.}
\newblock {\em The Neuroscientist : a review journal bringing neurobiology,
  neurology and psychiatry}, 14(5):487--502.

\bibitem[\protect\astroncite{Xu et~al.}{2007}]{Xu2007}
Xu, W., Huang, X., Takagaki, K., and Wu, J.-y. (2007).
\newblock {Compression and reflection of visually evoked cortical waves.}
\newblock {\em Neuron}, 55(1):119--29.

\end{thebibliography}

\end{document}